\documentclass[11pt,dvips]{article}
\usepackage{epsfig,times} 

\usepackage{picinpar}
\usepackage{wrapfig}
\usepackage{floatflt}
\makeatletter
\renewcommand{\section}{\@startsection{section}{1}{0in}
	{0.4\baselineskip}{0.1\baselineskip}{\Large\bf}}
\renewcommand{\subsection}{\@startsection{subsection}{2}{0in}
	{0.25\baselineskip}{-\baselineskip}{\large\bf}}
\renewcommand{\subsubsection}{\@startsection{subsubsection}{3}{0in}
	{0.1\baselineskip}{-\baselineskip}{\normalsize\bf}}
\makeatother
\begin{document}

%
%  Session and Paper Code:
\thispagestyle{myheadings}
%
%  ***INSTRUCTIONS:***  Replace `OG 9.9.9' in the command argument below
%			with your assigned session and paper code:
\markright{SH 3.2.41}
\begin{center}
%
%  ***INSTRUCTIONS:***  Replace `Instructions for Preparation of Manuscript'
%			with your paper's title:
{\LARGE \bf Cosmic Ray Sun Shadow in Soudan 2 Underground Muon Flux}
\end{center}

%  Author List:
\begin{center}
%
%  ***INSTRUCTIONS:***  Replace authors and addresses below with your own:
%
{\bf W.W.M.~Allison,$^3$, G.J.~Alner$^4$, D.S.~Ayres$^1$,
W.L.~Barrett,$^6$, C.~Bode,$^2$, P.M.~Border$^2$, C.B.~Brooks$^3$, 
J.H.~Cobb$^3$, R.J.~Cotton$^4$,
H.~Courant$^2$, D.M.~Demuth$^2$, T.H.~Fields$^1$, H.R.~Gallagher$^3$, C.~Garcia-Garcia$^4$,
M.C.~Goodman$^1$, R.~Gran$^2$, T.~Joffe-Minor$^1$, T.~Kafka$^5$, S.M.S.~Kasahara$^2$,
W.~Leeson$^5$, P.J.~Litchfield$^4$, N.P.~Longley$^2$, W.A.~Mann$^5$, 
M.L.~Marshak$^2$, R.H.~Milburn$^5$,
W.H.~Miller$^2$, L.~Mualem$^2$, A.~Napier$^5$, W.P.~Oliver$^5$, G.F.~Pearce$^4$,
E.A.~Peterson$^2$, D.A.~Petyt$^4$, L.E.~Price$^1$, K.~Ruddick$^2$, M.~Sanchez$^5$,
J.~Schneps$^5$, M.H.~Schub$^2$, R.~Seidlein$^1$, A.~Stassinakis$^3$, J.L.~Thron$^1$,
V.~Vassiliev$^2$, G.~Villaume$^2$, S.~Wakely$^2$, N.~West$^3$, D.~Wall$^5$\\
(The Soudan 2 Collaboration)}\\
{\it $^{1}$Argonne National Laboratory, Argonne, IL 60439, USA\\
$^{2}$University of Minnesota, Minneapolis, MN 55455, USA\\
$^3$Department of Physics, University of Oxford, Oxford OX1 3RH, UK\\
$^4$Rutherford Appleton Laboratory, Chilton, Didcot, Oxfordshire OX11 0QX, UK\\
$^5$Tufts University, Medford MA 02155, USA\\
$^6$Western Washington University, Bellingham, WA 98225, USA}
\end{center}

%  Abstract:
\begin{center}
{\large \bf Abstract\\}
\end{center}
\vspace{-0.5ex}

The absorption of cosmic rays by the sun produces a shadow at
the earth. The angular offset and broadening of the shadow are determined
by the magnitude and structure of the
interplanetary magnetic field (IPMF) in the inner solar system. We
report the first measurement of the solar cosmic ray shadow by detection of
deep underground muon flux in observations made during the entire
ten-year interval 1989 to 1998. The sun shadow varies significantly
during this time, with a $3.3\sigma$ shadow observed during the
years 1995 to 1998. 
%

%  Leave this line skip in place:
\vspace{1ex}

\section{Introduction}
\label{intro.sec}
The interplanetary magnetic field (IPMF) is produced by the trapping of the
sun's magnetic field in the solar wind. The
Archimedean spiral model for the IPMF, first described by Parker (1963), suggests
that away from the immediate vicinity of the sun, the IPMF field lines lie preferentially
in the region of the solar equatorial plane (nearly the ecliptic plane), varying in direction
from radial near the sun to $\approx 40^{\circ}$ from radial
at 1 AU.  This curvature is caused caused by the sun's rotation. 
Although the IPMF has been studied for many years, most measurements at $\le 1$ AU
are from satellite data recorded
about 1 percent of the distance from the earth to the sun.

The entire IPMF in the inner region of the ecliptic plane ($0 \le r \le 1$ AU) can be measured in
the aggregate sense by measuring the cosmic ray shadow of the sun. This measurement
may be used, in conjunction with satellite magnetic field measurements near the earth,
to test the validity of field models. Several such measurements have been made using
surface air-shower arrays. For many of these measurements, however, the
combination of poor detector resolution, high cosmic ray energies and short
observing times makes clear inferences about the IPMF difficult
(Alexandreas, 1991; Borione, 1994; Merck, 1996). 
The Tibet Air Shower Array is the only detector to
directly relate cosmic ray shadow characteristics to IPMF variability (Amenomori, 1996).

The Soudan 2 deep underground, iron, tracking calorimeter provides a new mode--deep 
underground muons rather than air showers--for measuring the cosmic
ray shadow of the sun. This detector has been described in detail elsewhere (Allison, 1996).
Most important for this discussion is that Soudan 2 has observed a clear cosmic
ray shadow of the moon (Cobb, 1999) with a statistical significance of $5\sigma$ and an
apparent Gaussian point spread function with $\sigma_r = 0.29^{\circ}$ (including
angular dispersion due to detector resolution and alignment, geomagnetic deflection,
shower and muon production and the $0.26^{\circ}$ radius of the moon). Differences between
this lunar shadow and the simultaneously observed solar shadow may be ascribed
to the magnetic field in the sun's immediate vicinity and the IPMF.

The expected offset and broadening of the solar cosmic ray shadow can be estimated
from the satellite magnetic field measurements using the Parker model.
Because the field lines radiate out from the sun and are only constrained to the
solar equatorial (or ecliptic) planes by a $\sin \theta$ factor, flux conservation
requires that the radial field $B_r$ (the major field component between the
sun and the earth) varies approximately as $1/r^2$, where $r$ is the 
distance from the sun. 
Using geocentric solar ecliptic coordinates (GSE), in the ecliptic plane near the
sun, $B_x$ dominates ($x$ points from the earth to the sun) and $B_z$ and $B_y$
are small ($z$ points normal to the ecliptic, positive in the direction 
closer to the earth's north pole; $y$ completes a right-handed, Cartesian coordinate system).
The Archimedean spiral model ($\theta = ar$, where a is a constant),
implies that for small $\theta$
($\theta \approx \sin \theta$), $\frac{B_y}{|B|}$, increases proportional to $r$. Thus,
$B_y$ varies as $1/r$ between the sun and the earth.
$B_z$ in the ecliptic plane remains a small fraction of $B_x$ out to large
distances, with a radial dependence likely between $1/r$ and $1/r^2$.  

In the impulse approximation, $\theta = \frac{0.3 \int B_tdx}{p}$, where $\theta$
is the deflection angle in mr, $p$ is the particle momentum in TeV/c, $B_t$ is the
magnetic field transverse to the particle trajectory in T and $x$ is the path length
in m. The sun-earth distance is $1.5 \times 10^{11}$ m or $\approx 215$ solar radii.
The shadow offset is determined by the mean tranverse field and the mean momentum. The
shadow broadening is determined by the mean and rms transverse fields, the mean
momentum and the rms momentum dispersion. The average daily mean (rms daily mean) values
for $B_y$ and $B_z$ measured by satellites for the entire 1989-1998 interval
are $B_y = -0.11$ nT (3.43 nT) and $B_z = 0.11$ nT (1.85 nT)
(OmniWeb, 1999). Monte Carlo studies
suggest that the mean momentum for cosmic ray primaries from the direction
of the sun producing muons at Soudan 2 is 20 TeV/c and the rms momentum dispersion
is of similar magnitude. Table 1 below shows the expected offset and broadening of the
solar cosmic ray shadow in the north-south (due to $B_y$) and east-west (due to $B_z$)
planes, using a $1/r$ dependence for $B_y$ and both a $1/r$ and $1/r^2$ dependence
for $B_z$. The table separately shows the expected broadening due to magnetic field
variation and momentum dispersion, although these two effects are of
approximately equal magnitude under the assumptions that $\Delta p \approx p$
and $\Delta B \approx B$. The rightmost column shows the expected combined effect calculated
as a quadrature. The calculations used to determine the table entries are as follows:
Assume $B_{obs}$ is the $y$ or $z$ component of $B$ measured near the earth.
For a $1/r$ dependence, $p\theta = B_{obs} \times 1.5 \times 10^{11} m \times 0.3
\times f$, where $f$ is approximately $(\ln~215 - \ln~10) = 2.4$, since 215 is the
number of solar radii in 1 AU and 10 is the approximate distance from the sun in solar
radii at which the Parker model becomes valid. For a $1/r^2$ dependence, 
$p\theta = B_{obs} \times 3.2 \times 10^{12} m \times (0.3)$, a quantity which is
9 times larger than the 1/r value. It is clear from the values in Table 1
that observation of a shadow distinguishes between $1/r$ and $1/r^2$
dependences for $B_z$. Even for $1/r$ dependences, observation of a shadow in
the Soudan 2 data is considerably more probable during intervals in which the
magnetic field is smaller than its average value over the 1989-1998 decade.

\vspace{1ex}
Table 1. Offset and broadening angles expected for solar shadow based on 
IPMF measurements and the Parker model.
\vspace{0.1in}

\begin{tabular}{|lcccc|}
\hline
 &Offset Angle&Broadening Angle&Broadening Angle&Broadening Angle\\
 & &$B$ Variation&$p$ Variation&Overall\\
 &(Degrees)&(Degrees)&(Degrees)&(Degrees)\\
\hline
$B_y$ (1/r dependence) &0.08&1.1&1.1&1.6\\
$B_z$ (1/r dependence) &0.08&0.57&0.57&0.8\\
$B_z$ ($1/r^2$ dependence &0.72&5.1&5.1&7.2\\
\hline
\end{tabular}
\vspace{1ex}

\section{Data Collection and Analysis}
\label{data.sec}

The data sample, collection procedure and muon track analysis used for the sun shadow
data are similar to those used for a moon shadow analysis. The background,
that is the number of events expected in the absence of a shadow, has been
estimated differently. The background algorithm used for the sun shadow is
to generate a 100 times real sample ensemble of pseudo-events, using random
combinations of arrival times and arrival directions in detector coordinates
of real events. The pseudo-events are then analyzed in the same way as real
events and pseudo-event distributions divided by 100 are then compared to
real event distributions.

Fig. 1(a) shows a plot of the angular density of muons,
$(1/\pi)(dN_{\mu}/d\theta^2)$ vs. $\theta$, the angular distance between the
muon direction and the calculated position of the center of the sun. 
The background distribution (not shown) indicates that in the absence of
a shadow this plot should be flat. The real events, however, 
show a deficit at small angles.  This deficit is both shallower and wider than
the Soudan 2 muon shadow for the moon observed during the same interval,
as is expected because of the effect of the IPMF. The significance of the shadow is tested 
by fitting the real event distribution to the form
\begin{equation}
\frac{dN_{\mu}}{d\theta^2} = \pi \lambda (1 - \pi(R_s^2/2\sigma^2)exp(
-\theta^2/2\sigma^2))
\end{equation}
where the unshadowed density $\lambda = 526.8\pm0.3$ is determined from the background
and $R_s$ (the apparent radius of the sun) and $\sigma$ are fitted parameters.  
$\sigma$ folds together all offset and broadening effects including deflections
due to the solar magnetic field and the IPMF, the finite angular size of the sun,
geomagnetic deflections, shower and muon production effects, multiple Coulomb
scattering and the the angular resolution and directional alignment of the detector.
The best fit parameters for the entire data sample are
$R_s = 0.174^{\circ}\pm0.026^{\circ}$--significantly less than the 
geometric size of the sun--and $\sigma = 0.59^{\circ}  
\mbox{}^{+0.17^{\circ}}_{-0.10^{\circ}}$--significantly more than the
detector angular resolution. The chance probability for the 
improvement in the $\chi^2$ from 
74.2 (80 df) for no fit to 63.5 (78 df) with a two-parameter fit is $4.8 \times 10^{-3}$.

The results of a two-dimensional analysis for the same data sample is shown in Fig. 1(b).
The center of the sun is at the center of the plot. The horizontal and vertical axes
are displacements in degrees measured parallel and perpendicular to the ecliptic
plane using $0.02^{\circ}$ by $0.02^{\circ}$ bins. The bin contents for both
real and background events have then been smoothed with a $\sigma = 0.59^{\circ}$ kernel.
The plot in Fig. 1(b) is a contour map of $z$ in units of standard deviations of
the difference between real and background event distributions. The sun shadow is
clearly visible, with a maximum depth of $z = -3.94$ at $0.18^{\circ}$ ecliptic longitude
and $-0.44^{\circ}$ ecliptic latitude from the sun.

The 10-year data collection interval spans much of an 11-year solar cycle
during which solar magnetic field and sunspot activity 
peaked in 1989-1991 and was at a minimal level in 1996.
Maps similar to the one in Fig. 1(b) are shown in Fig. 1(c) for 1989 to 1994 and
Fig 1(d) for 1995 to 1998, each with approximately half of the total number of
events. The average daily satellite-measured IPMF (sunspot number) was 
6.7 nT (104) for the former
interval and 4.8 nT (28) for the latter one. For the interval of high
field and sunspot activity, Fig. 1(c)
shows no evidence for a distinct shadow. The shadow is clearly seen in Fig. 1(d), however, 
for the years with low field and sunspot activity. An analysis similar to the one similar to the
one described above for Fig. 1(a) yields a chance probability for the shadow in Fig. 1(d)
of $5.2 \times 10^{-4}$ or $3.3 \sigma$ as a one-tailed Gaussian probability. For
Fig. 1(d), $R_s = 0.21^{\circ} \pm 0.03^{\circ}$, again less than the geometric size
of the sun.

\section{Conclusions}
\label{conclusions.sec}

We have observed the shadow of the sun in deep underground muons with a chance
probability of $4.8 \times 10^{-3}$ during the years 1989 to 1998. The existence
of a shadow implies that the IPMF normal to the ecliptic plane varies no more
steeply than $1/r$, where $r$ is the distance from the sun. An examination of 
approximately half of the data, that collected during 1989-1994, 
an interval of high magnetic field and
sunspot activity, shows no evidence for a distinct shadow. The other half of the data sample
collected during 1995-1998 shows a clear shadow
with a chance probability equivalent to $3.3 \sigma$.   

%
%  References: (DO NOT ALTER NEXT 4 LINES)
\vspace{1ex}
\begin{center}
{\Large\bf References}
\end{center}
%
%  ***INSTRUCTIONS:***  Enter your references alphabetically following the format
%			of the example citations below.
Alexandreas,~D.E.,~{\it et al.}, 1991 Phys. Rev. \textbf {D43}, 1735.\\
Allison,~W.W.M.,~{\it et al.}, 1996 Nucl. Inst. and Meth. \textbf {A376},
 377 and \textbf {A381}, 385.\\
Amenomori,~M.,~{\it et al.}, 1996 Astrophys. J. \textbf {464}, 954.\\
Borione,~A., ~{\it et al.}, 1994 Phys. Rev. \textbf {D49}, 1171.\\
Cobb,~J.H.,~{\it et al.}, 1999 ICRC Paper SH 3.2.42.\\ 
Merck,~M., ~{\it et al.}, 1996 Astropart. Phys. \textbf {5}, 379.\\
OmniWeb, 1999 Natl. Space Sci. Data Center, NASA Goddard Space Flight Center,
Greenbelt MD, USA.\\ 
Parker,~E.N., 1963 ${\it Interplanetary~Dynamical~Processes}$, Interscience, New York.
\begin{figwindow}[1,r,%
{\mbox{\epsfig{file=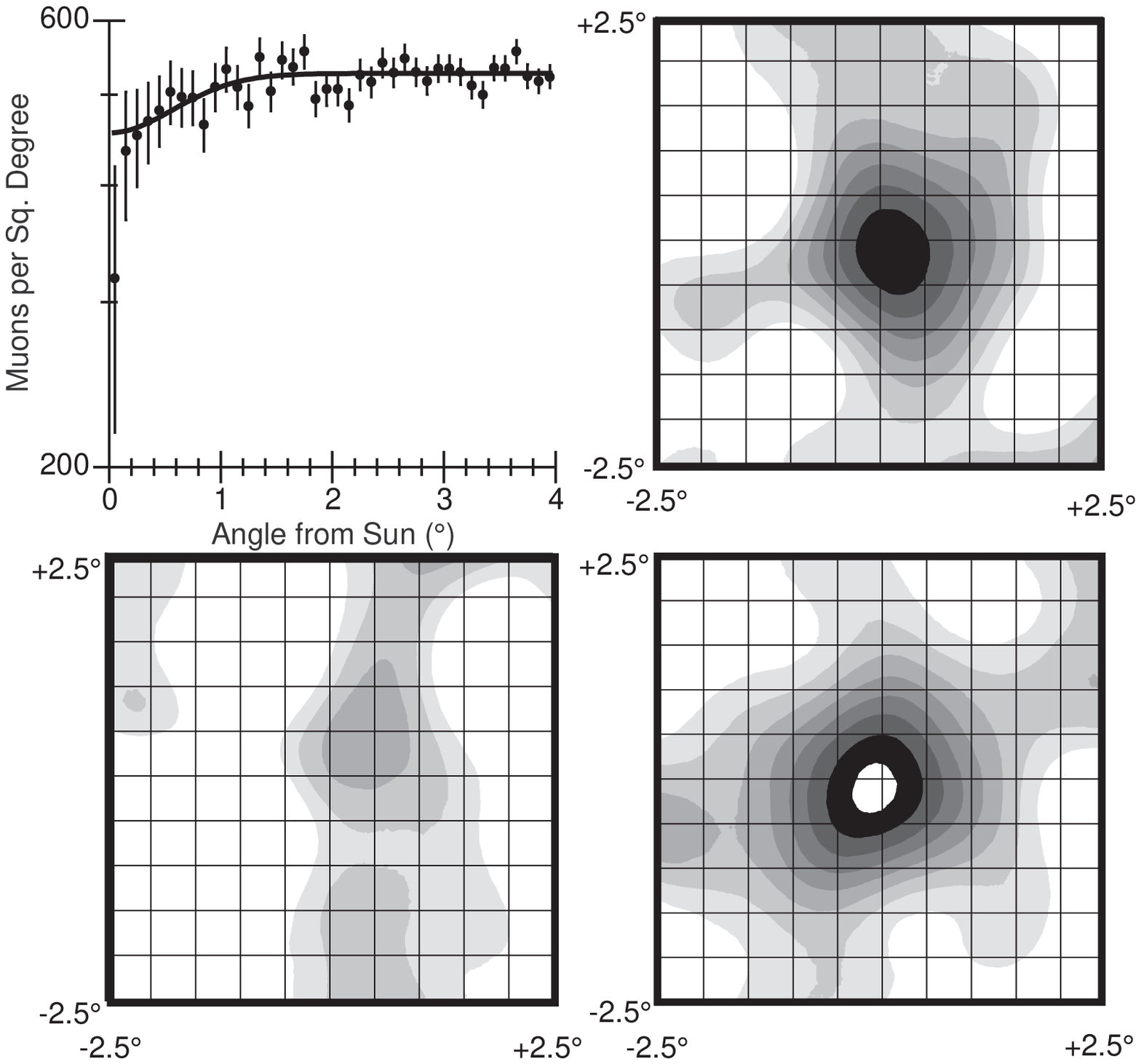,width=6.6in}}},%
{(a) (upper left) The angular density of muons, $(1/\pi)(dN_{\mu}/d\theta^2)$ 
vs. $\theta$, the angular distance between the
muon direction and the calculated position of the center of the sun. 
The line is a fit to the data using Eq. 1. 
(b) (upper right) Contour map of the normalized deviations, $Z$, for a $\pm2.5^{\circ}
\times \pm2.5^{\circ}$ region centered on the sun with a rebinning kernel
$\sigma_k = 0.59^{\circ}$. The contour lines are spaced by $\Delta Z = 0.5$
and are shown only where $Z \le 0.0$.
(c) (lower left) Same as (b) for the years 1989 to 1994. (d) (lower right)
Same as (b) for the years 1995 to 1998, except that the unshaded region in the
center of the plot indicates $-4.5 < Z < -4.0$. }]
\end{figwindow}

\end{document}